\documentclass[submission]{eptcs}
\usepackage{amsmath}
\usepackage{amssymb}
\usepackage{latexsym}
\usepackage{url}
\usepackage{color}
\usepackage{verbatim}
\usepackage{graphicx}
\usepackage{tikz}
\usepackage{mathpartir}
\usepackage{bussproofs}
\usepackage{space}

\newtheorem{theorem}{Theorem}
{\bfseries}{\rmfamily}
{\bfseries}{\rmfamily}
\newtheorem{PRO}[theorem]{Proposition}{\bfseries}{\itshape}
{\bfseries}{\rmfamily}
\newtheorem{LEM}[theorem]{Lemma}{\bfseries}{\itshape}
{\bfseries}{\itshape}
{\bfseries}{\itshape}
{\bfseries}{\rmfamily}

\newcommand{\NI}{\noindent}

\newcommand{\CAL}[1]{\mathcal{#1}}

\newcommand{\ENCan}[1]{\langle #1 \rangle}

\newcommand{\ASET}[1]{\{#1\}}
\newcommand{\PAR}{\ | \ }

\newcommand{\GAssert}{\mathcal{G}}  
\newcommand{\GA}{\GAssert}
\newcommand{\LAssert}{\CAL{L}}  
\newcommand{\LA}{\LAssert}

\newcommand{\USort}{U}
\newcommand{\SSort}{S}

\newcommand{\participant}[1]{\ensuremath{\mathtt{#1}}}
\newcommand{\q}{\ensuremath{\participant{q}}}
\newcommand{\p}{\ensuremath{\participant{p}}}

\newcommand{\BRA}[6]{{#1}[\p,\q]?\{#2(#3)\ENCan{#4}.#5 \}_{#6}}
\newcommand{\BRAF}[8]{{#1}[#7,#8]?\{#2(#3)\ENCan{#4}.#5 \}_{#6}}

\newcommand{\casedo}{\mapsto}
\newcommand{\SEL}[8]{#1[\p,\q]! \{ #2\casedo#3\ENCan{#4}(#5)\ENCan{#6};{#7}\}_{#8}}
\newcommand{\SELF}[9]{#1[#9]! \{ #2\casedo#3\ENCan{#4}(#5)\ENCan{#6};{#7}\}_{#8}}

\newcommand{\ABOUTPUT}[2]{\overline{#1}(#2)}

\newcommand{\inact}{\mathbf{0}}

\newcommand{\ProjA}[2]{#1 \upharpoonright {#2}}

\newcommand{\AOUT}[5]{\overline{#1}[#2](#3).#5}
\newcommand{\AIN}[5]{#1[#2](#3).#5}

\newcommand{\TRANS}[1]{\xrightarrow{#1}}

{\end{array}%
 \right.}

\newcommand{\keyword}[1]{\textsf{\upshape #1}}

\newcommand{\truek}{\keyword{true}}
\newcommand{\falsek}{\keyword{false}}

\newcommand{\kend}{\keyword{end}}

\newcommand{\proves}{\vdash}
\newcommand{\has}{\triangleright}

\newcommand{\SUBS}[2]{[{#1}/{#2}]}

\newcommand{\lsumOut}[6]{
#1! \{ #2(#3) \{#4\}\ENCan{#5}.#6 \}
}

\newcommand{\lsumIn}[6]{
#1? \{ #2(#3) \{#4\}\ENCan{#5}.#6 \}}

\newcommand{\newNP}[4]%
{\mathsf{new}\ #1 \ \mathsf{with} \ #2[#3]\ \mathsf{in} \ #4}

\newcommand{\newNPmo}[3]%
{\mathsf{new}\ #1 \ \mathsf{with} \ #2\ \mathsf{in} \ #3}

\newcommand{\newNPD}[5]%
{\mathsf{new}\ #1 \ \mathsf{with} \ #2[#3], #5\ \mathsf{in} \ #4}

\newcommand{\COMMENT}[1]{}

\newcommand{\hastype}{\triangleright}

\newcommand{\effected}[2]{#1\,\mathtt{after}\,#2}

\newcommand{\PRG}{P}

\newcommand{\PRGQ}{Q}

\newcommand{\AEnv}{\mathcal{C}}

\newcommand{\newNPmoPR}[3]%
{\mathsf{new}\ #1 \ \mathsf{with} \ [#2]_{\sigma'}\ \mathsf{in} \ #3}

\newcommand{\AEnvD}[2]{(#1,#2)}
\newcommand{\AEnvEx}[2]{\AEnvD{\AEnvL}{\AEnvR}}
\newcommand{\AEnvL}{A_{\AEnv}}
\newcommand{\AEnvR}{A_{\Lock}}

\newcommand{\IEnv}{\mathcal{I}}

\newcommand{\Lock}{\blacktriangledown}

\newcommand{\LockedInput}[6]
{#1[#2, #3]  ?\Lock \{ #4(#5). #6\}}

\newcommand{\myparagraph}[1]{\paragraph{\textbf{#1}}}

\newcommand{\OPEQ}{\ensuremath{=}}
\newcommand{\CODE}[1]{{\small \texttt{#1}}}

\newcommand{\DTYPE}[1]{\mathtt{#1}}
\newcommand{\ROLEVAR}[2]{\participant{#1}.{\participant{#2}}}
\newcommand{\OPINC}{\CODE{++}}
\newcommand{\GSEND}[2]
	{\ensuremath{\participant{#1} \rightarrow \participant{#2} :}}
\newcommand{\MSGlxS}[3]
	{\ensuremath{\MSGlx{#1}{#2\!:\!#3}}}  
\newcommand{\MSGlx}[2]
	{\ensuremath{#1 (#2)}}          

\newcommand{\PA}{~|~}

\newcommand{\fno}[1]{\mathtt{free}(#1)}

\newcommand{\Hfa}[2]{[#1]#2}

\newcommand{\impl}{\Rightarrow}

\newcommand{\out}{\ABOUTPUT}
\newcommand{\provlog}[2]{\vdash_{bool} #2}
\newcommand{\substi}[2]{\{#1 \slash #2\}}

\newcommand{\Nat}{\mathtt{Nat}}
\newcommand{\Gam}{\Gamma}

\newcommand{\embed}[2]{\|#1\|^{#2}}
\newcommand{\sembed}[2]{\|#1\|^{#2},\mathtt{S}}

\newcommand{\shfl}{\rtimes}
\newcommand{\Shfl}[1]{\huge\mbox{$\shfl$}\normalsize_{#1}}

\newcommand{\ertyp}[1]{\mathbf{Er}(#1)}
\newcommand{\Formu}[1]{\mathtt{Inter}(#1)}

\newcommand{\pembed}[1]{\|#1\|_{\mathtt{p}}}

\newcommand{\LASS}[1]
	{\ensuremath{\langle #1 \rangle}}  
\newcommand{\RASS}[1]
	{\ensuremath{\{ #1 \}}}            

\providecommand{\event}{PLACES 2013} 
\usepackage{breakurl}             

\title{Embedding Session Types in HML\thanks{This work has been
    partially sponsored by the project Leverhulme Trust award Tracing
    Networks, Ocean Observatories Initiative and EPSRC EP/K011715/1, EP/K034413/1 and EP/G015635/1.}} \author{Laura Bocchi
  \institute{Imperial College, London}  \email{l.bocchi@imperial.ac.uk} \and Romain Demangeon
  \institute{Imperial College, London} \email{rdemang@gmail.com} }

\def\titlerunning{Embedding Session Types in HML}
\def\authorrunning{L. Bocchi \& R. Demangeon}

\begin{document}
\maketitle

\begin{abstract}
Recent work on the enhancement of multiparty session types with
logical annotations enable the effective verification of properties on
(1) the structure of the conversations, (2) the sorts of the messages,
and (3) the actual values exchanged. In~\cite{BDY12} we extend this
work to enable the specification and verification of mutual effects of
multiple cross-session interactions. Here we give a sound and complete
embedding into the Hennessy-Milner logic to justify the expressiveness
of the approach in~\cite{BDY12} and to provide it with a logical
background that will enable us to compare it with similar approaches.
\end{abstract}

\section{Introduction}
\label{sect:introduction}
The Hennessy-Milner Logic (HML) is an expressive modal logic with a
strong semantic characterisation~\cite{HennessyM:alglawfndac} that enables the
specification of arbitrary behavioural properties of processes. 
Recent work on the enhancement of multiparty session types with logical annotations~\cite{BHTY10,BDY12}
addressed key challenges for logical specifications of processes, which were unexplored in the context of HML, 
such as the tractability of specifications of multiparty choreographies.

The work in~\cite{BHTY10,BDY12}  is based on multiparty session types~\cite{HYC08,BHTY10,DBLP:conf/tgc/CoppoD08} and 
inherits the same top-down approach. The key idea is that conversations are built as the composition of units
of design called {\em sessions} which are specified from a global
perspective (i.e., as a global type). Each global type is then
{\em projected} into one local type for each participant, making the responsibilities of each endpoint
explicit. This approach enables: (1) the effective verification of properties such as session
fidelity, progress, and error freedom, and (2) the modular local verification (i.e., of each
principal) of global properties of multiparty interactions.

The direct use of HML for the same purpose would require to start from endpoint specifications 
and then to check their mutual consistency, and would not offer the same tractability.
Starting from global assertions, instead, results in significant concision, while still enjoying generality in the 
modelling and verification of choreographies. 

By enhancing multiparty session types with logical annotations, \cite{BHTY10} enables the effective verification of properties on the actual values exchanged, other than the properties on the sorts of the messages guaranteed by~\cite{HYC08,DBLP:conf/tgc/CoppoD08}. 
For instance, global type $G$ in (\ref{eq:type}) describes, following a similar syntax to~\cite{DBLP:conf/tgc/CoppoD08}, a conversation where role $\participant{S}$ sends role $\participant{C}$ an integer and then continues as specified by global type $G'$. 
Following~\cite{BHTY10}, assertion $\GA$ in (\ref{eq:type}) can be obtained by annotating global type $G$; assertion $\GA$ further prescribes that the exchanged value, say $y$, must be greater than $10$. Note that $y$ is bound in $\GA'$ and the fact $\{y>10\}$ can be relied on in the subsequent interactions occurring in $\GA'$. 
\begin{equation}\label{eq:type}
\small
G=	\GSEND{S}{C}{(\DTYPE{int})}.G' \qquad 
\GA = 	\GSEND{S}{C}{(y:\DTYPE{int})}\{y>10\}.\GA'
\end{equation}
 
In \cite{BDY12} we extended  \cite{BHTY10} with the capability to refer to {\em virtual states} local to each network principal, hence 
expressing not only properties confined to the single multiparty sessions, but also stateful specifications incorporating mutual effects of multiple sessions run by a principal.  
\begin{equation}\label{eq:assertion}
\small
\begin{array}{l}
	\GSEND{S}{C}\MSGlxS{}{y}{\DTYPE{int}}\{y>10\land y
	= \ROLEVAR{S}{x}\}\ENCan{\ROLEVAR{S}{x} \OPINC}
\end{array}
\end{equation}
Consider now the protocol in (\ref{eq:assertion}). The
description of this simple distributed application implies
behavioural constraints of greater depth than the basic communication
actions. The (sender-side) \emph{predicate and effect} for the
interaction step,
$\{y>10\land y \OPEQ \ROLEVAR{S}{x}\}\ENCan{\ROLEVAR{S}{x} \OPINC}$, asserts
that the message $y$ sent to each client must equal the current value
of \ROLEVAR{S}{x}, a state variable $x$ allocated to
the \emph{principal} serving as \participant{S};
and that the local effect of sending this message is to increment
\ROLEVAR{S}{x}.
In this way, \participant{S} is specified to send incremental values across \emph{consecutive} sessions.
The resulting global specifications are called {\em multiparty stateful assertions} (MPSAs), and model the skeletal structure of the interactions of a session, the constraints on the exchanged messages and on the branches to be followed, and the {\em effects} of each interaction on the virtual state.

In order to obtain a clear understanding of the status of the logical methodology proposed in~\cite{BDY12}, it is useful to relate 
its notion of assertion to a more standard approach in process logic. This enables us to integrate different methods catering for different concerns, for which we may need a common logical basis. 
In this paper we consider the HML with predicates in \cite{BHTY10,DBLP:conf/icalp/BergerHY08}, and we justify the relevance of the stateful logical 
layer of~\cite{BDY12} by embedding the behaviours of each role in a session -- i.e., the projections of MPSAs -- into a HML formula. 
In this way, the required predicates will hold if a process and its state perform reductions and updates matching those of the specification. 
\begin{equation}\label{eq:assertion2}
\small\begin{array}{l}
\forall
y:\DTYPE{Nat}, \Hfa{s_{\mathtt{C}}(y)} {(y
= \ROLEVAR{S}{x} \land \Hfa{\ROLEVAR{S}{x}\OPINC} {\mathtt{true}})}
\end{array}
\end{equation}
(\ref{eq:assertion2}) is the formula corresponding to the behaviour of $\participant{S}$ in (\ref{eq:assertion}) on channel $s$, where 
$\Hfa{\ell}{\phi}$ means ``if a process and its state perform the action $\ell$, the resulting pair satisfies $\phi$''.
Communications and state updates are treated as actions of a
labelled transition system. 

We explain how specifications handling several roles in several sessions can be
soundly and completely embedded, through the use of an
\emph{interleaving} of formulae, exploring all the possible orders in
which the actions coming from different sessions can be performed, and
ensuring that predicates are always satisfied.


\section{HML Embedding}
\label{asec:hml}

\myparagraph{Logical layer} We propose an embedding of our analysis
into Hennessy Milner Logic (HML), together with soundness and
completeness results. The analysis in~\cite{BDY12} be seen as the
superposition of two analyses: a session type system and a logical
layer. The former ensures that a process is able to perform some
visible actions described by the specification and can be
mechanically, yet tediously encoded in HML, for instance, by using a
``surely/then'' modality \cite{DBLP:conf/icalp/BergerHY08}.  Our
contribution focuses on the embedding of the latter, namely on
\emph{predicate safety}, ensuring that stateful predicates will be
satisfied.  As consequence, the completeness result we propose
(Proposition~\ref{prop:hml:cscomp}), states that if a process abides
to the session-type component $L$ of a local assertion $\LA$ (obtained by
erasing all predicates in $\LA$) and satisfies the logical encoding of
$\LA$, then it is provable against $\LA$.

\paragraph{MPSAs} We focus here on local assertions, 
each referring to a specific role and deriving, via projection, from a
global assertion as in~\cite{BDY12} -- e.g., as (\ref{eq:assertion}).
Local assertions are defined by the grammar below and are ranged over
by $\LA$.

 {\small
$$
\begin{array}{lll} 
\LA & ::= & 
\lsumOut{\p}{l_i}{x_i:\USort_i}{A_i}{E_i}{\LA_i}_{i \in I}  
\mid
\lsumIn{\p}{l_i}{x_i:\USort_i}{A_i}{E_i}{\LA_i}_{i \in I}
\\
&& \mu t\{y:A'\}(x:\SSort).\LA:A 
\mid
t(y:A') 
\mid
\kend  
\end{array}
$$
}

Selection $\lsumOut{\p}{l_i}{x_i:\USort_i}{A_i}{E_i}{\LA_i}_{i \in I}$
models an interaction where the role sends $\p$ a branch label $l_i$
and a message $x_i$ of sort $U_i$ (e.g., $\mathtt{int}$,
$\mathtt{bool}$, etc., and local assertion for delegation) and continues as $\LA_i$, with being $A_{i}$
predicates\footnote{As in~\cite{BHTY10,BDY12} we assume that the
  validity of closed formulae is decidable.} and $E_{i}$ state
updates. Branching $\lsumIn{\p}{l_i}{x_i:\USort_i}{A_i}{E_i}{\LA_i}_{i \in I}$ is dual to selection. 
We use guarded recursion defining a recursion parameter $x$ initially set equal to a value satisfying the initialisation predicate $A'$, 
where $y$ is the free variable of $A'$, and with $A$ being an invariant predicate. 
Recursive call $t(y:A')$ instantiates a new iteration of $t$ where the recursion parameter takes a value satisfying $A'$, with $y$ free variable of $A'$. 

\cite{BDY12} uses local assertions as a basis for the verification of a processes, ranged over by $P$. 
\newcommand{\ptp}[1]{\mathtt{#1}}

\[\begin{array}{lllll}
   \PRG &\;::=\;&
     \inact
     \mid 
  \AOUT{u}{\ptp{n}}{y}{A}{\PRG}
  \mid
  \AIN{u}{\ptp{i}}{y}{A}{\PRG}
\mid
  \SEL{k}{e_i}{l_i}{e'_i}{x_i}{E_i}{\PRG_i}{i\in I}
\mid
  \BRA{k}{l_i}{x_i}{E_i}{\PRG_i}{i \in I}  
\\[1mm]
&&  \PRG \PAR \PRGQ ~~
\mid
  (\mu X(x).\PRG) \langle{e}\rangle
\mid
  X\langle{e}\rangle
   \end{array}\]
A process can be an idle process $\inact$, a session request/accept, 
a guarded command~\cite{Dijkstra:1975:GCN:360933.360975}, a branching, a parallel composition of processes, a recursive definition and invocation. 
Session request $\overline{u}[\ptp{n}](y).P$ multicasts a 
request to each session accept process ${u}[\ptp{i}](y).P$ (with $i\in \{2,..,n\}$) by synchronisation through a shared name $u$ and continuing as $P$. Guarded command and branching processes represent communications through an established session $k$.
Guarded command $\SEL{k}{e_i}{l_i}{e'_i}{x_i}{E_i}{P_i}{i\in I}$ acts as role $\p$ in session $k$ and sends role $\q$ 
one of the labels $l_i$. The choice of the label is determined by boolean expressions $e_i$, assuming $\lor_{i\in I}e_i=\truek$ and $i\not= j$ implies $e_i\land e_j=\falsek$.  Each label $l_i$ is sent with the corresponding expression $e'_i$ which specifies the value for $x_i$, assuming $e'_i$ and $x_i$ have the same type. Branching $\BRA{k}{l_i}{x_i}{E_i}{P_i}{i\in I}$ plays role $\q$ in session $k$ and is ready to receive from $\p$ one of the labels $l_i$ and a value for the corresponding $x_i$, then behaves as $P_i$ after instantiating $x_i$ with the received value. In guarded command (resp. branching), the local state of the sender (resp. receiver) is updated according to update $E_i$; in both processes each $x_i$ binds its occurrences in $P_i$ and $E_i$.

The judgements are of the form $\AEnv ; \Gam \vdash P \has  \Delta$ where: 
\begin{itemize}
\item $\AEnv$ is the assertion environment that is the set of preconditions built, during the verification, as the 
incremental conjunction of the predicates occurring in the branchings, 
\item $\Gam$ determines which types of sessions can be initiated by a process by mapping shared channels to 
global assertions (e.g.,  if $\Gamma(a)=\mathtt{I}(\GA)$ then $P$ can be invited to join a session specified by $\GA$), 
\item $\Delta$ is the session environment mapping sessions that $P$ has joined, say $s[\p]$, to local types. 
\end{itemize}
We write omit $\Gam$ (resp. $\AEnv$) in the judgment when it is the empty mapping (resp. $\truek$ precondition). 

\paragraph{HML} 

Here, the behaviour prescribed for $P$ is modelled using the standard
HML with the first-order predicates as in
\cite{DBLP:conf/icalp/BergerHY08}. We use the same type of predicate
$A$ as in MPSAs. We associate this HML with a LTS where actions $\ell$
model communications and state updates.
$$
\begin{array}{rll}
\ell & ::= & s[\p,\q](x) \mid \out{s[\p,\q]}{x} \mid E
\end{array}
$$
Namely, $s[\p,\q](x)$ is an input action, $\out{s[\p,\q]}{x}$ is an output action, and $E$ is a state update. We let states to be ranged over by $\sigma,\sigma',\ldots$ and we write $\sigma' = \effected{\sigma}{\ell}$ for the state $\sigma'$ obtained by updating $\sigma$ as prescribed by $E$. 
$P,\sigma \TRANS{\ell}P',\sigma'$ if either:  (a) $\ell$ is an input or output action, $P \TRANS{\ell} P'$ and $\sigma' = \sigma$, or (b) $\ell$ is an update action, $P = P'$, and $\sigma' = \effected{\sigma}{\ell}$. 

We use $\phi$ to denote HML-formulae, which are built from predicates, implications, universal
quantifiers, conjunctions and \emph{must} modalities. The logic used
in this \emph{safety embedding} is positive: if we remove the
implication symbol, there is no negation, no existential quantifier,
no disjunction and no may modality. Additionally, the implication will
always appear as $A \impl \phi$ meaning that modalities never appear
in the negative side.
$$
\begin{array}{rll}
 \phi  ::=  \mathtt{true} \mid \phi \land \phi \mid \phi \impl \phi \mid
 \Hfa{\ell}{\phi} \mid A \mid \forall x:S.\phi  
\end{array}
$$









The satisfactions rules (Figure~\ref{fig:hml_rules}) are fairly
standard. For a pair $P,\sigma$ to satisfy a predicate $A$, written $P,\sigma\models A$, $A$ has to
hold with respect to $\sigma$, denoted by $\sigma \provlog{}{A}$, meaning
that $\sigma(A)$ is a tautology for the boolean logic.

\begin{figure}
\[\begin{array}{ccc}
\inferrule{P,\sigma \models \phi_1 \and P,\sigma \models \phi_2}{P,\sigma
  \models \phi_1 \land \phi_2}
\quad \inferrule{~}{P,\sigma \models \mathtt{true}}
\quad \inferrule{\text{ if }P,\sigma \models \phi_1\text{ then
  }P,\sigma\models \phi_2}{P,\sigma \models \phi_1 \impl \phi_2}\\
  \inferrule{\text{For all } P',\sigma'\text{ s.t. } P,\sigma
  \TRANS{\ell} P',\sigma' ,  P',\sigma'\models \phi }{P \models
  \Hfa{\ell}{\phi}}
\quad \inferrule{\sigma \provlog{\emptyset}{A}}{P,\sigma \models A}
\quad \inferrule{\text{For all values } v \text{ of type }T, P,\sigma
  \models \phi\SUBS{v}{x}}{P,\sigma \models \forall x:T.\phi}
\end{array}\]
\caption{Logical rules}
\label{fig:hml_rules}
\end{figure}


The embedding of local types we propose is parameterised with a session
channel $s[\p]$. Predicates appearing in input prefixes are embedded
as premises in implications, as in (\ref{emb2}), and predicates in output prefixes have to
be satisfied, as in  (\ref{emb1}), yielding:

\vspace{-0.3cm}
{
\begin{equation} \label{emb1}\embed{\lsumOut{\q}{l_i}{x_i:S_i}{A_i}{E_i}{\LA_i}
_{i \in I}}{s[\p]} =  \bigwedge_{i \in I} \forall x_i:S_i,
  \Hfa{\out{s[\p,\q]}{x_i}}{(A_i \land
    \Hfa{E_i}{\embed{\LA_i}{s[\p]}})} 
\end{equation}
\vspace{-0.3cm}
\begin{equation} \label{emb2}
\embed{\lsumIn{\q}{l_j}{x_j:S_j}{A_j}{E_j}{\LA_j}_{j \in
    J}}{s[\p]}  =  \bigwedge_{j \in J} \forall x_j:S_j,
\Hfa{s[\q,\p](x_j)}{(A_j \impl
  \embed{\LA_j}{s[\p]})} 
\end{equation}}

\noindent The embedding of selection (\ref{emb1}), is a conjunction of the formulae
corresponding to the branches: for each value sent on the session
channel, predicates should be satisfied and, if the state is updated,
the embedding of the continuation should hold. For branching types (\ref{emb2}),
the assertion is used as an hypothesis and no update appears.

\section{Soundness} 

For the sake of clarity, we divide our proofs into two parts, one
proving \emph{simple} preciseness, that is soundness and completeness
when the specification is a single session type, the other proves the
\emph{full} completeness, for any specification. This corresponds to
the two challenges we tackle in our approach: the translation of a type
into a formula, and the handling of the possible interleaving of
concurrent types.

\paragraph{Simple Preciseness}

We postpone the introduction of interleavings to focus on proving our
result for single types, obtaining a \emph{simple} preciseness result.

The following lemma states that a process cannot perform an action on
a channel that does not appear in its type, that a process that does
not perform any action does not change the set of formulae it satisfies, 
that satisfaction of assertions is stable by reduction and
that validity of satisfaction judgements is stable by well-typed
substitutions.

\begin{LEM}
\label{lemma:hml:typsaf}
If $\AEnv;\Gam \proves P \has  \Delta$ and $s[\p] \notin \Delta \cup
\Gam$, then there is no $P'$ s.t. $P,\sigma \TRANS{\ell_s}
P',\sigma$ for any action $\ell_s$ of the form $s![\p,\q](x)$ or $s![\q,\p](x)$.

Similarly, if $a:\mathtt{I}(\mathcal{G}) \notin \Gam$, there are no
$P'$ and $s[\p]$ such that $P,\sigma \TRANS{a(s[\p])} P',\sigma$.

~\label{lemma:hml:comptriv}
If $P_1,\sigma \models \phi$ and $P_2$ cannot make any action, then
$P_1 \PA P_2, \sigma \models \phi$.

~\label{lemma:hml:stabass}
If $P,\sigma \models A$ and $P \TRANS{\ell} P'$, then $P',\sigma
\models A$.

~\label{lemma:hml:satsub}
If $P,\sigma \models \phi$ and $x:S,v:S$ are not bound in $P,\sigma$
and $\phi$, then $P\SUBS{v}{x},\sigma \models \phi\SUBS{v}{x}$.
\end{LEM}

\paragraph{Proof}
  By induction on $\phi$, as our processes and formulas abide a
  Barendregt convention, the case $\forall y.\phi$ is easy as $y \neq
  x$ and $y \neq v$. The only interesting cases are assertion and must modality:
\begin{itemize}
\item Case $\phi = A$. The logic rules notifies that $\sigma(A)$ is a
  tautology, so any instantiation of its free variable should be
  so. Thus $\sigma(A)\substi{v}{x}$ is a tautology and any process (in
  particular $P\substi{v}{x}$) and the state $\sigma$ form a pair that
  satisfies it.
\item Case $\phi = \Hfa{\alpha}{\phi'}$. We prove, by induction on the
  reduction rules, that if $P \TRANS{\alpha} P'$, then $P\substi{v}{x}
  \TRANS{\alpha\substi{v}{x}} P'\substi{v}{x}$ and use the induction
  hypothesis.
\end{itemize}

We state, thanks to the previous lemmas, the following `simple'
soundness for simple local types, that is for $\Delta$ with one single
local type:
\begin{PRO}[Simple Soundness]
~\label{prop:hml:ssound} If $\AEnv \proves \PRG \hastype
  s[\p]:\LA $, then $(P,\sigma) \models \AEnv \impl
  \embed{\LA}{s[\p]}$.
\end{PRO}


In order to state simple completeness we define unasserted types. Unasserted types are built from:
\begin{mathpar}
\begin{array}{llllll}
L ::= & \hspace{2mm}\p?\ASET{l_i(U_i).L_i}_{i \in I} 
 & \mid \p!\ASET{l_i(U_i).L_i}_{i \in I} 
& \mid \mu t . L \mid t \mid \kend \end{array}
\end{mathpar}
An unasserted local type can be obtained from an asserted local type using an erasing operator. 
The erasing operator $\ertyp{\LA}$ is defined by the removal of every
assertion, update and variable from $\LA$. Unasserted typing rules for the
judgements $\proves P \hastype \Delta$ are easily deduced from the asserted ones.


\begin{PRO}[Simple Completeness]
~\label{prop:hml:scompl} For all $\LA$, if $ \proves P \has
  s[p] :\ertyp{\LA}$ and $P ,\sigma \models \AEnv \impl
  \embed{\LA}{s[\p]} $ then $ \AEnv  \proves \PRG \hastype
  s[p]:\LA $.
\end{PRO}

\paragraph{Proof}
By induction on the typing judgement $ \proves P \has  s[\p]:\ertyp{\LA}$:
\begin{itemize}
\item Case \emph{branching}. We have $\LA =
  \lsumIn{\p_0}{l_i}{x_i:U_i}{A_i}{E_i}{\LA_i}_{i \in I}$. Let $i \in
  I$ and suppose $\AEnv$ holds. We have from the hypothesis $\proves P
  \hastype \p_0?\ASET{l_i(U_i).L_i}_{i \in I}$. The unasserted typing
  rules give that $P = \BRAF{s}{l_i}{x_i}{E_i}{\PRG_i}{i \in I}{\p_0}{\p}$,
  and $\proves P_i \hastype s[\p]:L_i $.  We know that $P , \sigma
  \models \AEnv \implies \embed{\LA}{s}$, which
  is: \begin{mathpar} P,\sigma \models \bigwedge_{i \in I} \forall
    x_i.\Hfa{\out{s}{x_i}}{(A_i \implies
      \Hfa{E_i}{\embed{\LA_i}{s}} \land (A_i \land
      \AEnv )}\end{mathpar}
  rules, that $P$ can perform $s(x_i)$ to $P_i,\sigma \models (A_i
  \implies \Hfa{E_i}{}\embed{\LA_i}{s[\p]} \land A_i )$. We see that $\sigma$ can
  perform $E_i$ to $\effected{\sigma}{E_i}$, meaning that we have $P_i
  , \effected{\sigma}{E_i} \models (A_i \implies
  \embed{\LA_i}{s[\p],\IEnv})$, we use the induction hypothesis to get
  $\AEnv \land A_i \proves P_i \hastype \LA_i$. To sum up, for all
  $i$, $\AEnv, A_i\proves P_i \hastype s[\p]:\LA_i$. We use the proof
  rule for branching to prove $\AEnv \proves P \hastype s[p]:\LA$.
\item Case \emph{selection}. We have $\LA =
  \lsumOut{\p_0}{l_i}{U_i}{A_i}{E_i}{\LA_i}_{i \in I}$. Suppose $\AEnv$
  holds and $\sigma \models \IEnv$. We have from the hypothesis
  $\proves P \hastype s[\p]:\p_0?\ASET{l_i(U_i).L_i}_{i \in I}$. The
  unasserted typing rules give $P =
  \SELF{s}{e_j}{l_j}{e'_j}{x_j}{E_j}{\PRG_j}{j\in J}{\p,\p_0}$, and
  $\proves P_j \hastype s[\p]:L_j $.  We know that $P,\sigma \models
  \AEnv \implies \embed{\LA}{s}$, which is $P \models \bigwedge_{i \in
    I} \forall x_i.\Hfa{\out{s}{x_i}}{A_i \land \embed{\LA_i}{s[\p]}
    \land (A_i \land \AEnv )}$. In particular, as $\AEnv$ holds, $P
  \models \Hfa{\out{s[\p,\p_0]}{x_j}}{A_j \land \embed{\LA_j}{s[\p]} \land
    (A_j ) )}$ We know from the shape of $P$, given above, and the
  reduction rules, that $P$ can perform $\out{s[\p,\p_0]}{x_j}$ to $P_j
  \models (A_j \land \embed{\LA_j}{s})$, meaning that $A_j$
  holds. Also, $\sigma$ can perform $E_j$ to
  $\effected{\sigma}{E_j}$. To sum up, we have $\AEnv 
  \implies A_j$, $P_j \models \AEnv \implies
  \embed{\LA_j}{s[\p]}$ and $\proves P_j \hastype s[\p]:\LA_j$,
  we use the induction
  hypothesis to get $\AEnv \proves P_j \hastype:\LA_j$ and this allows
  us to use the proof rule for selection to prove $\AEnv \proves P
  \hastype s[\p]:\LA$.
\item Case \emph{parallel}. No assertion appear in the parallel rule
  and we can use Lemmas~\ref{lemma:hml:typsaf}.1
  and~\ref{lemma:hml:comptriv}.2 to state that exactly one side of the
  parallel composition satisfies the formula (along with the same
  state $\sigma$). As a consequence, we use the induction hypothesis
  twice and conclude.
\item Case \emph{end}. $\LA = \kend$, so this case is trivial.
\end{itemize}


\paragraph{Full preciseness}

Full preciseness is done using the previous simple results, and
additional lemmas handling interleavings.

To obtain soundness for typing judgements involving specifications,
we have to introduce \emph{interleavings} of formulae, treating the fact
that one process can play several roles in several sessions. As a
simple example both $s[\p_1,\p_2]?(x).k![\q_1,\q_2]$ $\langle
10\rangle$ and $k![\q_1,\q_2]\langle 10\rangle.s[\p_1,\p_2]?(x)$
can be typed with $s[\p_2]:\p_1?(x:\Nat).\kend,$
$k[\q_1]:\q_2!(y:\Nat).\kend.$

Interleaving is not a new operator \emph{per se} and can be seen as
syntactic sugar, describing shuffling of must modalities. The main
rule for interleaving is: $ \Hfa{\ell_1}{\phi_1} \shfl
\Hfa{\ell_2}{\phi_2} = \Hfa{\ell_1}{(\phi_1 \shfl
  \Hfa{\ell_2}{\phi_2})} \land \Hfa{\ell_2}{(\Hfa{\ell_1}{\phi_1}
  \land \phi_2)}$.  When interleaving two or more formulae containing
modalities, we obtain a conjunction of formulae, each one representing
a different way of organising all modalities in a way that preserves their
initial orders. Informally, the interleaving of $\Hfa{1}{\Hfa{2}{}}$
and $\Hfa{A}{\Hfa{B}{}}$ is
{$\Hfa{1}{\Hfa{2}{\Hfa{A}{\Hfa{B}{}}}} \land
  \Hfa{A}{\Hfa{B}{\Hfa{1}{\Hfa{2}{}}}} \land
  \Hfa{1}{\Hfa{A}{\Hfa{2}{\Hfa{B}{}}}} \land
  \Hfa{A}{\Hfa{1}{\Hfa{B}{\Hfa{2}{}}}} \land
  \Hfa{1}{\Hfa{A}{\Hfa{B}{\Hfa{2}{}}}} \land
  \Hfa{A}{\Hfa{1}{\Hfa{2}{\Hfa{B}{}}}} $}.

The full rules for interleaving are given in Figure~\ref{fig:int}.

\begin{figure}
\[\begin{array}{ccc}
 \Hfa{\ell_1}{\phi_1} \shfl
\Hfa{\ell_2}{\phi_2}   =  \Hfa{\ell_1}{(\phi_1 \shfl
  \Hfa{\ell_2}{\phi_2})} \land \Hfa{\ell_2}{(\Hfa{\ell_1}{\phi_1}
  \land \phi_2)}\\
\Hfa{\ell_1}{\phi_1} \shfl (\phi_{2,1} \land \phi_{2,2})  = 
\Hfa{\ell_1}{(\phi_1 \shfl \phi_{2,1})} \land \Hfa{\ell_1}{(\phi_1
  \shfl \phi_{2,2})}\\
\phi \shfl \mathtt{true}  =  \phi\\
 \phi \shfl (\phi_1 \land \phi_2)  =  (\phi \shfl \phi_1) \land
(\phi \shfl \phi_2)\\
(\phi_1 \land \phi_2) \shfl \phi  =  (\phi_1 \shfl \phi) \land
(\phi_2 \shfl \phi)\\
\forall x:T.\phi_1 \shfl \phi_2\\
(A \impl \phi_1) \shfl \phi_2  =  A \impl (\phi_1 \shfl
\phi_2)
\end{array}\]
\label{fig:int}
\caption{Rules for interleaving}
\end{figure}



We encode a pair $\Delta,\Gamma$ into a complex formula
$\Formu{\Delta,\Gamma}$, defined as the interleaving of the formulae
obtained by encoding the local types of $\Delta$ on their
corresponding channels and the formulae corresponding to $\Gamma$,
built as follows: for each channel $a:\mathtt{I}(\mathcal{G})$, if
some $s[\p]$ is received on $a$, the resulting process should satisfy
the encoding on $s[\p]$ of the projection of $\mathcal{G}$ on $\p$:
\[\Formu{s_1[\p_1],\dots,s_n[\p_n];a_1:\mathtt{I}(\mathcal{G}_1),
    \dots,a_m:\mathtt{I}(\mathcal{G}_m)}
= \embed{T_1}{s_1[\p_1]}
  \shfl \dots \shfl \embed{T_n}{s_n[\p_n]} \shfl 
\phi_1 \shfl \dots \shfl \phi_m
  \]
where 
$\phi_i=\forall   s'_i.\forall\p'_i.\Hfa{a_i(s'_i[\p_i])}
  {\embed{\ProjA{\mathcal{G}_i}{\p'_i}}{s'_i[\p'_i]}} $.

\begin{LEM}[Shuffling correctness]
~\label{lemma:hml:shcorr}

If $P_1 \models \phi_1$ and $P_2
\models \phi_2$ and if $\fno{\phi_1} \cap \fno{P_2} = \fno{\phi_2}
\cap \fno{P_1} = \fno{P_1} \cap \fno{P_2} = \fno{\phi_1}\cap
\fno{\phi_2} = \emptyset$, then $P_1 \PA P_2 \models \phi_1 \shfl
\phi_2$.

Conversely, if $P_1 \PA P_2 \models \phi_1 \shfl \phi_2$, and 
$\fno{\phi_1} \cap \fno{P_2} = \fno{\phi_2} \cap \fno{P_1} = \fno{P_1}
\cap \fno{P_2} = \fno{\phi_1}\cap \fno{\phi_2} = \emptyset$, then
$\fno{\phi_1} \subseteq \fno{P_1}$ and $\fno{\phi_2} \subseteq
\fno{P_2}$.

\end{LEM}

\paragraph{Proof}
We proceed by double structural induction over the pair
$(\phi_1,\phi_2)$.
\begin{itemize}
\item The most interesting case is when both formula are modalities:
  $\phi_1 = \Hfa{\alpha_1}{\phi'_1}$ and $\phi_2 =
  \Hfa{\alpha_2}{\phi'_2}$. The formula $\phi_1 \shfl \phi_2$ is
  $\Hfa{\alpha_1}{(\phi'_1 \shfl \phi_2)} \land
  \Hfa{\alpha_2}{(\phi'_1 \shfl \phi_2)}$. We prove that $P_1 \PA
  P_2$ satisfies the first formula (the other part is similar). First
  the condition of $\fno{P_2} \cap \fno{\phi_1}$ ensures that there is
  no $P'_2$ such that $P_2 \TRANS{\alpha_1} P'_2$. As a consequence, if
  $P_1 \PA P_2 \TRANS{\alpha_1} P'$, it means that $P_1
  \TRANS{\alpha_1} P'_1$. By hypothesis, $P'_1 \models \phi'_1$ and we
  use the induction hypothesis to get $P'_1 \PA P_2 \models (\phi'_1
  \shfl \phi_2)$.
\item The other cases are treated by destructing one construct, following
  the definition, and using the induction hypothesis.
\end{itemize}

\begin{LEM}[Description of free names]
~\label{lemma:hml:desfr}
If $\AEnv,\Gam \proves P\has \Delta$ then $\fno{P} \subseteq
\fno{\Delta}\cup\fno{\Gamma}$

\end{LEM}

Easily done by induction on the typing judgement.

\begin{LEM}[Nature of an interleaving]
~\label{lemma:hml:nature}

Let $\Delta = \{ s_k[\p_k]: \q_k\begin{array}{l}! \\ ?
\end{array}
\{l_i(x_i:U_i) \{A_i\}\langle E_i\rangle.T_{k,i}\}_{i \in I} \}_k $ and $\Gamma = \{
a_j:\mathtt{I}(\mathcal{G}_j) \}_j$ be well-formed, then the formula
$\Formu{\Delta,\Gamma}$ is equivalent to a formula guarded by several
$\forall$ operators guarding a conjunction of formulae, each one
starting with a modality, and this modalities are in bijection with
the pairs of $(s_k[\begin{array}{l}\p_k,\q_k \\ \q_k,\p_k
  \end{array}
],l_{k,i})$ and $(a_j,\emptyset)$.
\end{LEM}
%



\paragraph{Proof}
By induction on the typing judgment:
\begin{itemize}
\item Case \emph{selection}. In this case we have $P = \BRAF{s}{l_i}{x_i}{E_i}{\PRG_i}{{i \in
      I}}{\p}{\p_0}$ and $ \Delta = \Delta', s[\p]:
  \p_0?\ASET{l_i(x_i:U_i)\RASS{A_i}\LASS{E_i}.\LA_i}_{i \in I}$. 
  We use Lemma~\ref{lemma:hml:nature} to state the formula
  we are trying to validate using $P$ is a conjunction on several
  formulas, all beginning with a different modality from the pairs
  $(s_k[\p_k],l_{k,i})$ and $(a_j,\emptyset)$ . As $P$ is only able
  to perform an action $s[\p,\p_0]?$, all formulas starting with a
  modality associated to a different name are automatically satisfied,
  and we have to prove that for each $i$:
  \begin{mathpar}\PRG ,\sigma \models \AEnv
  \implies \sembed{T_i}{s[\p]} \shfl \Shfl{s_k[\p_k]:T_k \in \Delta'}
  \sembed{T_k}{s_k[\p_k],\IEnv} \shfl \Shfl{a_j:\mathcal{G}_j[\p_j]
    \in \Gam} \forall
  s_j.\Hfa{a_j(s_j[\p_j])}{\sembed{\mathcal{G}_j|_{p_j}}{s_j[\p_j],\IEnv}}
  \end{mathpar} 
  We conclude in a way similar to the one followed 
in the proof of Proposition~\ref{prop:hml:ssound}.
\item Case \emph{branching}. We have $P = \SELF{s}{e_i}{l_i}{e'_i}{x_i}{E_i}{\PRG_i}{i \in
      I}{\p,\p_0}$. We use Lemma~\ref{lemma:hml:nature} to state the formula
  we are trying to validate using $P$ is a conjunction on several
  formulas, all beginning with a different prefix. As $P$ is only able
  to perform an action $s[\p,\p_0]!$, all formulas starting with a
  different modality are automatically satisfied, and we have to prove
  We conclude using the proof of Proposition~\ref{prop:hml:ssound}.
\item Case \emph{session reception}. We have $P = a(s).P'$ and $\Gam =
  a:\mathcal{G}[\p],\Gam'$. We use Lemma~\ref{lemma:hml:nature} to
  state the formula we are trying to validate using $P$ is a
  conjunction on several formulas, all beginning with a different
  modality. As $P$ is only able to perform an action on $a$, all
  formulas starting with a modality associated to a different name are
  automatically satisfied, and we have to prove that $P$ satisfies
  $\forall
  s[\p],\Hfa{a(s)}{\Formu{\Gam';\Delta,s:\mathcal{G}|_{\p}}}$. As $P$
  is able to receive $s[\p]$ on $a$, we use the induction hypothesis
  to conclude.
\item Case \emph{parallel composition}. Easily done by using
  Lemmas~\ref{lemma:hml:shcorr} and~\ref{lemma:hml:desfr} and the fact
  that both $\Gam$ and $\Delta$ are split multiplicatively in the rule
  for parallel composition we use.
\item Case \emph{end} is trivial.
\end{itemize}

We extend the erasing operator to $\Delta$. Namely, $\ertyp{\Delta}$ maps $s[\p]$ to $\ertyp{\LA}$ iff $\Delta$ maps $s[\p]$ to $\LA$.
Our preciseness result is:

\begin{PRO}[Preciseness]
\label{prop:hml:cssound}
If $ \Gam \proves \PRG \hastype \Delta $, then: $\PRG ,\sigma \models
(\Formu{\Delta,\Gamma})$.  ~\label{prop:hml:cscomp} If $\proves \PRG
\hastype \ertyp{\Delta} $ and $\PRG ,\sigma \models
(\Formu{\Delta,\Gamma})$ then $ \Gam \proves \PRG \hastype \Delta $
\end{PRO}

By induction on the unasserted typing judgment, case branching and
selection are treated in a way similar to the proof of
Proposition~\ref{prop:hml:scompl}, parallel composition is done using
Lemmas~\ref{lemma:hml:shcorr} and~\ref{lemma:hml:desfr}.















\section{Refinements} 

\paragraph{Embedding to pure HML} We are actually able to embed a stateful
satisfaction relation $P,\sigma \models \phi$ into a satisfaction
relation $P' \models \phi'$ for a standard $\pi$-calculus with
first-order values, by encoding the store $\sigma$ into a
$\pi$-process:
$$
 \begin{array}{l}\pembed{x_1 \mapsto v_1, \dots, x_n \mapsto v_n}{}
  = \quad\quad \out{a_1}{v_1} \PA \dots \PA \out{a_n}{v_n} \PA
  \\ \quad !x_1(e).a_1(y_1)\dots
  a_n(y_n).(\out{a_1}{\mathtt{eval}(e\SUBS{y_1 \dots y_n}{x_1 \dots
      x_n})} \PA \out{a_2}{y_2} \PA \dots \PA \out{a_n}{y_n} ) \PA
  \dots \PA \\ \quad !x_n(e).a_1(y_1)\dots a_n(y_n).(\out{a_1}{y_1} \PA
  \dots \PA \out{a_{n-1}}{y_{n-1}} \PA
  \out{a_n}{\mathtt{eval}(e\SUBS{y_1 \dots y_n}{x_1 \dots x_n})})
\end{array}$$


\NI For each variable $x_i$ in the domain of the state $\sigma$, we add
an output prefix emitting its content on the channel $a_i$ and we add
a replicated module that waits for an update $e$ at $x_i$, then
capture the value of all variables of the current state, replace the
variable $x_i$ by evaluating $e$ by $\mathtt{eval}$, and then makes
available the other ones. Soundness and completeness allow us to state
that HML formulae for pairs state/process can be seen as pure HML
formulas on the $\pi$-processes.


The embedding for the formula is given by 
\[\begin{array}{lll}
\pembed{\Hfa{E}{\phi}} & = & \Hfa{\pembed{E}}{\pembed{\phi}}\\
\pembed{A} & = & \Hfa{\out{x_1}{v_1}}{\dots\Hfa{\out{x_n}{v_n}}
  {A\substi{v_1,\dots,v_n}{x_1,\dots,x_n}}}
  \end{array}\]
  where the state variables of $A$ are $x_1,\dots,x_n$.

\begin{PRO}[Preciseness]
~\label{prop:hml:puresound}
If $P,\sigma \models \phi$, then $\pembed{P} \PA \pembed{\sigma} \models
\pembed{\phi}$.

If $\pembed{P} \PA \pembed{\sigma} \models \pembed{\phi}$ then $P,\sigma
\models \phi$
\end{PRO}


\paragraph{Embedding Recursion}

Recursion can be encoded at the cost of much technical details. We add
to our HML syntax the recursion operators, $\mu X.\phi$ and $X$
(similar to the ones present in the
$\mu$-calculus~\cite{DBLP:journals/tcs/Dam94}).
The main difficulty lies in the interaction between interleaving and
recursion: loops coming from different sessions can be interleaved in
many different way, and the difficult task is to compute the finite
formula which is equivalent to this interleaving.
As a small example consider the following session environment
(interactions are replaced by integer labels): $s_1[\p_1]: \mu
X.1.2.X, s_2[\p_2]: \mu Y.3.4.Y$. The simplest HML formula describing
all possible interleavings is: 

{ $$\begin{array}{l} \mu A.([1]\mu B.([2]A \land [3]\mu
C.([4]B \land [2]([1]C \land [4].A) )) \land  \\
 ~[3]\mu D.([4].A \land
[1]\mu E.([2]D \land [4]([2]A \land [3]E))))
  \end{array}$$}


We use the following method to obtain a matching HML formula. We use a
translation through finite automata. Here is a sketch of the method,
which takes as arguments a set session environment $\Delta$:
\begin{enumerate}
\item Encode every session judgement $s_i[\p_i]:T_i$ of $\Delta$ into
  a formula $\phi_i$, using $\embed{\mu X.T}{s[\p]}=\mu
  X\embed{T}{s[\p]}$.
\item Translate every formula $\phi_i$ into a finite automata
  $\mathcal{A}_i$, one state corresponds to a syntactic point between
  two modalities or a $\mu X$, one transition correponds to either
  $\Hfa{\ell}{(A \land \Hfa{E}{\circ})}$ (output) or $\Hfa{\ell}{(A
    \impl \circ)}$ (input). Every automata is \emph{directed} with a
  source state corresponding to the head of the formula and leaf
  states corresponding to recursion variables (or end of protocols).
\item Compute the automata $\mathcal{A}$, the parallel composition of
  all the $\mathcal{A}_i$, which is still \emph{directed}.
\item Expand the automata $\mathcal{A}$, in order to obtain an
  equivalent branch automata, that is, an automata such that there is
  a root (the starting state) and transitions form a tree (back
  transitions are allowed but only on the same branch). This could be
  done by recursively replacing sub-automata with several copies of
  this sub-automata.
\item Translate back the automata into a formula, every state with
  more than two incoming transition is encoded as a recursion
  operator.
\end{enumerate}

On our example, we obtain the formulas $\mu X.[1][2].X$ and $\mu
Y.[3][4].Y$,
each one giving an automaton with 2 states (initial and between $[1]$
(resp. $[3]$) and $[2]$ (resp. $[4]$)).
Merging yields automata with $4$ states: the initial one, one after
$[1]$, one after $[3]$, one after both $[1]$ and $[3]$. These automata
are diamond-shaped (hence not tree-shaped).
Expansion yields an automaton with $7$ states, which is then translated
in the formula described above.
The preciseness proof relies on the fact that the operation described
in 3. and 4. give equivalent automata, and that two formulas
translated into two equivalent automata are equivalent for the HML
satisfaction relation.






\section{Conclusion}



Hennessy-Milner logic (HML)  is a natural and semantically complete logic for processes which can
immediately be applied to the distributed $\pi$-calculus in~\cite{BDY12}. 
The HML with hypothetical supposition can faithfully embed the safety aspect of stateful MPSAs: at the same time, the restricted expressive
power of MPSAs enables tractable dynamic and static validations.  The
underlying type structures and linkage among them through local state
is a major reason why local types enable both static and runtime
verification against rich specifications. 

The work \cite{CF10} investigates a relationship between a dual
intuitionistic linear logic and binary session types, and shows that
the former defines a proof system for a session calculus which can
automatically characterise and guarantee a session fidelity and global
progress. In \cite{DBLP:conf/csfw/ArapinisRR11}, the authors introduce
 a state layer in a $\pi$-caclulus, toward the validation of security
properties for protocols. The work
\cite{BCF11} further extends \cite{CF10} to the dependent type theory
in order to include processes that communicate data values in
functional languages.  
A recent work \cite{ICFP11} encodes dynamic
features in~\cite{DY11} in a dependently typed language for secure
distributed programming.  
None of the above works treat either virtual
states or logical specifications for interleaved multiparty sessions.

    \bibliographystyle{eptcs}
    \bibliography{session}

\begin{thebibliography}{10}
\providecommand{\bibitemdeclare}[2]{}
\providecommand{\surnamestart}{}
\providecommand{\surnameend}{}
\providecommand{\urlprefix}{Available at }
\providecommand{\url}[1]{\texttt{#1}}
\providecommand{\href}[2]{\texttt{#2}}
\providecommand{\urlalt}[2]{\href{#1}{#2}}
\providecommand{\doi}[1]{doi:\urlalt{http://dx.doi.org/#1}{#1}}
\providecommand{\bibinfo}[2]{#2}

\bibitemdeclare{inproceedings}{DBLP:conf/csfw/ArapinisRR11}
\bibitem{DBLP:conf/csfw/ArapinisRR11}
\bibinfo{author}{Myrto \surnamestart Arapinis\surnameend},
  \bibinfo{author}{Eike \surnamestart Ritter\surnameend} \&
  \bibinfo{author}{Mark~Dermot \surnamestart Ryan\surnameend}
  (\bibinfo{year}{2011}): \emph{\bibinfo{title}{StatVerif: Verification of
  Stateful Processes}}.
\newblock In: {\sl \bibinfo{booktitle}{CSF}}, \bibinfo{publisher}{IEEE Computer
  Society}, pp. \bibinfo{pages}{33--47}, \doi{10.1109/CSF.2011.10}.

\bibitemdeclare{inproceedings}{DBLP:conf/icalp/BergerHY08}
\bibitem{DBLP:conf/icalp/BergerHY08}
\bibinfo{author}{Martin \surnamestart Berger\surnameend},
  \bibinfo{author}{Kohei \surnamestart Honda\surnameend} \&
  \bibinfo{author}{Nobuko \surnamestart Yoshida\surnameend}
  (\bibinfo{year}{2008}): \emph{\bibinfo{title}{Completeness and Logical Full
  Abstraction in Modal Logics for Typed Mobile Processes}}.
\newblock In: {\sl \bibinfo{booktitle}{ICALP (2)}}, {\sl
  \bibinfo{series}{LNCS}} \bibinfo{volume}{5126},
  \bibinfo{publisher}{Springer}, pp. \bibinfo{pages}{99--111},
  \doi{10.1007/978-3-540-70583-3\_9}.

\bibitemdeclare{inproceedings}{BDY12}
\bibitem{BDY12}
\bibinfo{author}{Laura \surnamestart Bocchi\surnameend},
  \bibinfo{author}{Romain \surnamestart Demangeon\surnameend} \&
  \bibinfo{author}{Nobuko \surnamestart Yoshida\surnameend}
  (\bibinfo{year}{2013}): \emph{\bibinfo{title}{A Multiparty Multi-session
  Logic}}.
\newblock In: {\sl \bibinfo{booktitle}{TGC}}, {\sl \bibinfo{series}{LNCS}}
  \bibinfo{volume}{8191}, \bibinfo{publisher}{Springer}, pp.
  \bibinfo{pages}{97--111}, \doi{10.1007/978-3-642-41157-1\_7}.

\bibitemdeclare{inproceedings}{BHTY10}
\bibitem{BHTY10}
\bibinfo{author}{Laura \surnamestart Bocchi\surnameend}, \bibinfo{author}{Kohei
  \surnamestart Honda\surnameend}, \bibinfo{author}{Emilio \surnamestart
  Tuosto\surnameend} \& \bibinfo{author}{Nobuko \surnamestart
  Yoshida\surnameend} (\bibinfo{year}{2010}): \emph{\bibinfo{title}{A Theory of
  Design-by-Contract for Distributed Multiparty Interactions}}.
\newblock In: {\sl \bibinfo{booktitle}{CONCUR}}, {\sl \bibinfo{series}{LNCS}}
  \bibinfo{volume}{6269}, pp. \bibinfo{pages}{162--176},
  \doi{10.1007/978-3-642-15375-4\_12}.

\bibitemdeclare{inproceedings}{CF10}
\bibitem{CF10}
\bibinfo{author}{Luis \surnamestart Caires\surnameend} \&
  \bibinfo{author}{Frank \surnamestart Pfenning\surnameend}
  (\bibinfo{year}{2010}): \emph{\bibinfo{title}{Session Types as Intuitionistic
  Linear Propositions}}.
\newblock In: {\sl \bibinfo{booktitle}{CONCUR}}, {\sl \bibinfo{series}{LNCS}}
  \bibinfo{volume}{6269}, \bibinfo{publisher}{Springer}, pp.
  \bibinfo{pages}{222--236}, \doi{10.1007/978-3-642-15375-4\_16}.

\bibitemdeclare{inproceedings}{DBLP:conf/tgc/CoppoD08}
\bibitem{DBLP:conf/tgc/CoppoD08}
\bibinfo{author}{Mario \surnamestart Coppo\surnameend} \&
  \bibinfo{author}{Mariangiola \surnamestart Dezani-Ciancaglini\surnameend}
  (\bibinfo{year}{2008}): \emph{\bibinfo{title}{Structured Communications with
  Concurrent Constraints}}.
\newblock In: {\sl \bibinfo{booktitle}{TGC}}, pp. \bibinfo{pages}{104--125},
  \doi{10.1007/978-3-642-00945-7\_7}.

\bibitemdeclare{article}{DBLP:journals/tcs/Dam94}
\bibitem{DBLP:journals/tcs/Dam94}
\bibinfo{author}{Mads \surnamestart Dam\surnameend} (\bibinfo{year}{1994}):
  \emph{\bibinfo{title}{{CTL*} and {ECTL*} as Fragments of the Modal
  mu-Calculus}}.
\newblock {\sl \bibinfo{journal}{TCS}}
  \bibinfo{volume}{126}(\bibinfo{number}{1}), pp. \bibinfo{pages}{77--96},
  \doi{10.1016/0304-3975(94)90269-0}.

\bibitemdeclare{inproceedings}{DY11}
\bibitem{DY11}
\bibinfo{author}{Pierre-Malo \surnamestart Deni{\'e}lou\surnameend} \&
  \bibinfo{author}{Nobuko \surnamestart Yoshida\surnameend}
  (\bibinfo{year}{2011}): \emph{\bibinfo{title}{Dynamic Multirole Session
  Types}}.
\newblock In: {\sl \bibinfo{booktitle}{POPL}}, pp. \bibinfo{pages}{435--446},
  \doi{10.1145/1926385.1926435}.

\bibitemdeclare{article}{Dijkstra:1975:GCN:360933.360975}
\bibitem{Dijkstra:1975:GCN:360933.360975}
\bibinfo{author}{Edsger~W. \surnamestart Dijkstra\surnameend}
  (\bibinfo{year}{1975}): \emph{\bibinfo{title}{Guarded commands,
  nondeterminacy and formal derivation of programs}}.
\newblock {\sl \bibinfo{journal}{Commun. ACM}} \bibinfo{volume}{18}, pp.
  \bibinfo{pages}{453--457}, \doi{10.1145/360933.360975}.

\bibitemdeclare{article}{HennessyM:alglawfndac}
\bibitem{HennessyM:alglawfndac}
\bibinfo{author}{Matthew \surnamestart Hennessy\surnameend} \&
  \bibinfo{author}{Robin \surnamestart Milner\surnameend}
  (\bibinfo{year}{1985}): \emph{\bibinfo{title}{Algebraic {L}aws for
  {N}on-{D}eterminism and {C}oncurrency}}.
\newblock {\sl \bibinfo{journal}{JACM}}
  \bibinfo{volume}{32}(\bibinfo{number}{1}), pp. \bibinfo{pages}{137--161},
  \doi{10.1145/2455.2460}.

\bibitemdeclare{inproceedings}{HYC08}
\bibitem{HYC08}
\bibinfo{author}{Kohei \surnamestart Honda\surnameend}, \bibinfo{author}{Nobuko
  \surnamestart Yoshida\surnameend} \& \bibinfo{author}{Marco \surnamestart
  Carbone\surnameend} (\bibinfo{year}{2008}): \emph{\bibinfo{title}{{Multiparty
  Asynchronous Session Types}}}.
\newblock In: {\sl \bibinfo{booktitle}{POPL'08}}, \bibinfo{publisher}{ACM}, pp.
  \bibinfo{pages}{273--284}, \doi{10.1145/1328438.1328472}.

\bibitemdeclare{inproceedings}{ICFP11}
\bibitem{ICFP11}
\bibinfo{author}{Nikhil \surnamestart Swamy\surnameend}, \bibinfo{author}{Juan
  \surnamestart Chen\surnameend}, \bibinfo{author}{Cedric \surnamestart
  Fournet\surnameend}, \bibinfo{author}{Pierre-Yves \surnamestart
  Strub\surnameend}, \bibinfo{author}{Karthikeyan \surnamestart
  Bharagavan\surnameend} \& \bibinfo{author}{Jean \surnamestart
  Yang\surnameend} (\bibinfo{year}{2011}): \emph{\bibinfo{title}{Secure
  Distributed Programming with Value-Dependent Types}}.
\newblock In: {\sl \bibinfo{booktitle}{ICFP}}, \bibinfo{publisher}{ACM}, pp.
  \bibinfo{pages}{266--278}, \doi{10.1145/2034773.2034811}.

\bibitemdeclare{inproceedings}{BCF11}
\bibitem{BCF11}
\bibinfo{author}{Bernardo \surnamestart Toninho\surnameend},
  \bibinfo{author}{Luis \surnamestart Caires\surnameend} \&
  \bibinfo{author}{Frank \surnamestart Pfenning\surnameend}
  (\bibinfo{year}{2011}): \emph{\bibinfo{title}{Dependent Session Types via
  Intuitionistic Linear Type Theory}}.
\newblock In: {\sl \bibinfo{booktitle}{PPDP}}, \bibinfo{publisher}{ACM}, pp.
  \bibinfo{pages}{161--172}, \doi{10.1145/2003476.2003499}.

\end{thebibliography}
%


\end{document}